\documentclass[prl,amsmath,amssymb,twocolumn]{revtex4}
\usepackage{amsmath,graphicx}
\usepackage{epsfig}

\begin{document}

\title{Forced Imbibition - a  Tool for Determining
Laplace Pressure,\\ Drag Force and Slip Length in Capillary Filling Experiments
}

\author{D. I. Dimitrov$^{1,3}$, A. Milchev$^{1,2}$, and K. Binder$^1$ \\
[\baselineskip]
\it$^{(1)}$ {\it Institut f\"ur Physik, Johannes Gutenberg Universit\"at Mainz,
}\\
{\it Staudinger Weg 7, 55099 Mainz, Germany}\\
$^{(2)}$ {\it Instute for Chemical Physics, Bulgarian Academy of Sciences,} \\
{\it 1113 Sofia, Bulgaria} \\
$^{(3)}$ {\it University of Food Technologies, } {\it 4002 Plovdiv, Bulgaria} \\
}

\begin{abstract}
When a very thin capillary is inserted into a liquid, the liquid is sucked into
it: this imbibition process is controlled by a balance of capillary and drag
forces, which are hard  to quantify experimentally, in particularly  considering
flow on the nanoscale. By computer experiments using a generic coarse-grained
model, it is shown that an analysis of imbibition forced by a controllable
external pressure quantifies relevant physical parameter such as the Laplace
pressure, Darcy's permeability, effective pore radius, effective viscosity,
dynamic contact angle and slip length of the fluid flowing into the pore. In
determining all these parameters independently, the consistency of our analysis
of such forced imbibition processes is demonstrated.
\end{abstract}

\maketitle

Flowing fluids confined to pores with diametera on the $\mu m$ to $nm$ scale are
important in many contexts: oil recovery from porous rocks~\cite{1}; separation
processes in zeolithes~\cite{2}; nanofluidic devices such as liquids in
nanotubes~\cite{3}; nanolithography~\cite{4}, nanolubrication~\cite{5}, fluid
transport in living organisms~\cite{6} and many other applications~\cite{1}.
However, despite its importance for so many processes in physics, chemistry,
biology and technology, the flow of fluids into (and inside) nanoporous
materials often is not well understood: the effect of pore surface structure on
the flowing fluid~\cite{5,7,8} is difficult to assess, in terms of
hydrodynamics, the problem differs dramatically from the macroscopic fluid
dynamics~\cite{9,10}; and although very beautiful experiments have recently been
made (e.g.~\cite{11,12}), more information is needed for a complete description
of the relevant microfluidic process.

In the present work, we propose to use \emph{forced imbibition} with the
external pressure as a convenient control parameter to obtain a much more
diverse information on the parameters controlling flow into capillaries than
heretofore possible. Extending our recent study of imbibition at zero
pressure~\cite{13}, we concisely describe the theoretical basis for this new
concept, and provide a comprehensive test of the concept in terms of a "computer
experiment" on a generic model system (a fluid composed of Lennard-Jones
particles flowing into a tube with a perfectly crystalline (almost
\begin{figure}[htb]
\includegraphics[scale=1.5]{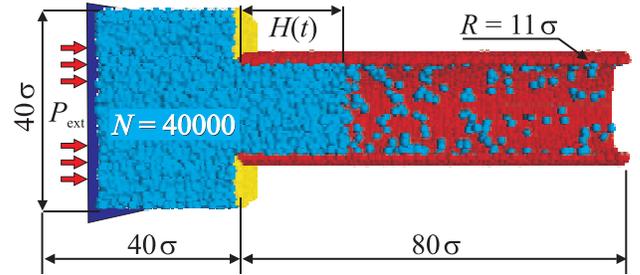}
\caption{A snapshot of the capillary during imbibition (cross section along the
axis of the cylindrical capillary oriented in $x$-direction). At the left side
there is a box, which initially has a cubic shape with linear dimensions of $40$
Lennard-Jones diameters $\sigma$, and contains $N=40000$ particles interacting
with the Lennard-Jones potential $U_{LJ}(r) = 4\epsilon[(\sigma/r)^{12} -
(\sigma/r)^6]$, $r$ being the distance between the particles, $\epsilon=1.4$ (in
temperature units, $k_B T=1$). The left wall of the box can move to maintain a
given pressure $P_{ext}$ in the reservoir box. The capillary has a radius
$R=11\sigma$ and length $80\sigma$. It ends at a hard wall on the right side
while it is open toward the left side, to connect with the reservoir box, which
has a circular opening in the right wall (shown in yellow).}
\label{snap}
\end{figure}
rigid~\cite{13}) wall and spherical cross-section, see Fig.~\ref{snap}. We also
provide a stringent test of our  description by having estimated all parameters
of the theory in independent earlier work ~\cite{13}), and hence there are no
adjustable parameters  whatsoever. We emphasize that our procedures and analysis
could be followed in experiments with real materials fully analogously.

We briefly summarize the pertinent theory. On a macroscopic scale the rise of a
fluid meniscus at height  $H(t)$ over the entrance of a capillary with time $t$
is described (at zero external pressure) by the well-known Lucas-Washburn
equation~\cite{14}
\begin{equation}\label{eq1}
H^2(t) =\left(\frac {\gamma _{LV}R \cos \theta}{2 \eta}\right)t + H^2_0.
\end{equation}
Here $\gamma_{LV}$ is the liquid-vapour surface tension of the liquid, $\eta$
the shear viscosity of the fluid, $R$ the pore radius, $\theta$ the contact
angle, and $H_0$ a constant (which accounts for the fact that Eq.(\ref{eq1})
holds only after some transient time when inertial effects have already
vanished).  Eq.(\ref{eq1}) follows when one balances the viscous drag force
$\frac{8\eta}{R^2} H(t) \frac{dH(t)}{dt}$ with the Laplace pressure $P_L =
2\gamma_{LV} \cos(\theta)/R$.

The applicability of Eq.(\ref{eq1}) for ultrathin pores has been rather
controversial~\cite{16,17,18}. This debate was clarified~\cite{13} by recalling
that on the nanoscale the slip length $\delta$~\cite{19,20} must not be
neglected. According to the definition of this length, the drag force under slip
flow conditions in a tube of radius $R$ and slip length $\delta$ is equal to the
drag force for a no-slip flow in a tube of effective radius $R+\delta$. Thus one
ends up with a modified Lucas-Washburn relationship:
\begin{equation}\label{eq2}
 \frac{2\gamma_{LV}\cos(\theta)}{R} + P_{ext} =
\frac{8\eta}{(R+\delta)^2}H(t)\frac{dH(t)}{dt}.
\end{equation}
On the left hand side of Eq.~\ref{eq2} we have now also included an external
pressure term $P_{ext}$. If one uses Darcy's permeability~\cite{21} $\kappa =
(R+\delta)^2/8$, Eq.~\ref{eq2} can be written in a form which does not depend on
the capillary radius anymore, introducing also the rate $v(P_{ext})=\frac{d
H^2(t)}{dt}$,
\begin{equation}\label{eq3}
 P_{L}+ P_{ext} = \frac{\eta}{\kappa}H(t)\frac{dH(t)}{dt} =
\frac{1}{2}v(P_{ext})\frac{\eta}{\kappa}.
\end{equation}
For constant $P_{ext}$, Eq.~\ref{eq3} is easily integrated to
\begin{equation}\label{eq4}
H^2(t) =\frac {2\kappa}{\eta}(P_{L}+ P_{ext})\;t + H^2_0.
\end{equation}

Eq.~\ref{eq3} shows that $v(P_{ext})$ varies linearly with $P_{ext}$, so
measuring this relationship yields {\em both} parameters $P_L$ and $\eta /
\kappa$. Instead of using the height $H(t)$ by observing the meniscus, one may
alternatively estimate $v(P_{ext})$ from the time variation of the mass of the
fluid inside the capillary (i.e. the total number of particles $N(t)\propto
H(t)$ which has entered the capillary up to the time $t$). In contrast, the
classical experiments on spontaneous imbibition of a fluid, where $P_{ext}=0$,
yield only the product $\kappa P_L$, and hence even if the fluid viscosity
$\eta$ is known, one cannot discern the effects due to the driving force
($\propto P_L$) and due to the drag force ($\propto \kappa$). Moreover,
Eq.~\ref{eq3} suggests the intriguing possibility of applying the present
concepts to the most general case of porous media~\cite{1}, {\em irrespective}
of the particular geometry and topology of the channels in such materials, but
this will not be followed up here.

We now present a test of the above concepts by a quantitative analysis of the
computer experiment outlined in Fig.~\ref{snap}. We assume also a Lennard-Jones
interaction (of strength $\epsilon_{WL}$ between the wall and the fluid
particles (see~\cite{13} for details on how the wall is atomistically modeled),
and study the cases of both nonwettable ($\epsilon_{WL}<0.65$) and wettable
\begin{figure}
\includegraphics[scale=0.55]{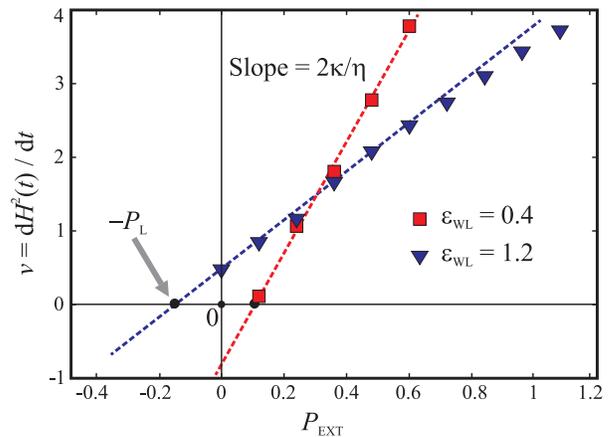}
\caption{Imbibition rate $v=dH^2(t)/dt$ plotted vs. external pressure $P_{ext}$,
for two choices of $\epsilon_{WL}$ (as indicated) that controls the wettability
of the capillary wall: for $\epsilon_{WL}=0.4$ the contact angle $\cos \theta
\approx -0.56$ (the fluid does not wet the wall); for $\epsilon_{WL}=1.2$ one
finds $\cos \theta \approx 0.92$, i.e. almost complete wetting~\cite{15}. The
dashed lines show fits to the observed $v(P_{ext})$ to straight lines, reducing
the range to $P_{ext} \le 0.6$. The slope of the straight lines  is equal to
$2\kappa/\eta$, the intersection point with the axis $v=0$ yields the Laplace
pressure $P_L$. Note that the Molecular Dynamics simulation applies the
velocity-Verlet algorithm with a timestep $\delta t = 0.01/\sqrt{48}$, choosing
the particle mass $m=1$~\cite{13} and units where $\sigma=1$ and $k_B T=1$.}
\label{v}
\end{figure}
($\epsilon_{WL}>0.65$) walls. Fig.~\ref{v} shows a plot of $v(P_{ext})$ vs.
$P_{ext}$. One can see that there is a broad regime where the variation of
$v(P_{ext})$ with $P_{ext}$ is indeed linear (deviations from linearity for
large $P_{ext}$ can be attributed to a slight increase of viscosity with
increasing fluid density at large pressures). Thus Fig.~\ref{v} demonstrates
that indeed  a rather precise estimation of both $P_L$ and $\kappa$ is possible.
This is important in many cases, e.g. nanocapillaries or porous media, where
neither $P_L$ nor $\kappa$ can be reliably predicted theoretically (because
information is missing, e.g. the effective channel radius $R$ or the (dynamic)
contact angle $\theta$ or the slip length $\delta$ may be unknown).

\begin{figure}
\includegraphics[scale=0.55]{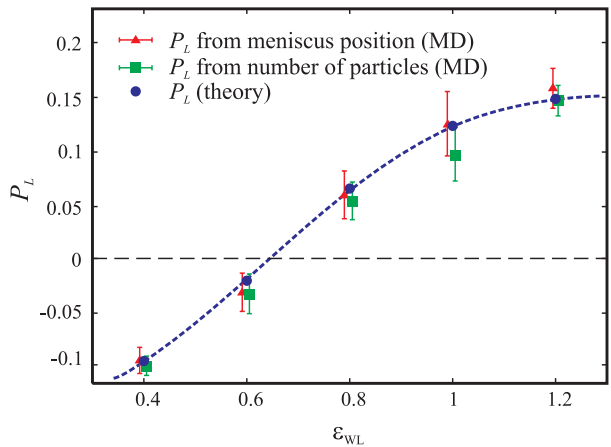}
\caption{Plot of the Laplace pressure $P_L$ against the strength $\epsilon_{WL}$
of the wall-liquid interaction. Triangles denote estimates from the meniscus
position $H(t)$, squares are derived from the number of particles $N(t)$ that
have entered the capillary. Full circles and a dashed line show the theoretical
prediction of $P_L=2\gamma\cos\theta/R$, cf. text. The estimate
$\gamma_{LV}=0.735 \pm 0.015$ was taken from \cite{13}. } \label{PL}
\end{figure}

From the Laplace pressure $P_L$ one can readily obtain information on the
contact angle (if interfacial tension $\gamma_{LV}$ and pore radius $R$ are
known). Fig.~\ref{PL} shows
the variation of $P_L$ with $\epsilon_{WL}$ in our model. By "measuring" the
contact angle $\theta$ dependence on $\epsilon_{WL}$ in a separate simulation,
as well as $\gamma_{LV}$, we can predict $P_L$ as $P_L=2\gamma_{LV}
cos\theta/R$, as noted above. Fig.~\ref{PL} shows that the agreement between
this prediction and the observations is excellent.

\begin{figure}
\includegraphics[scale=0.6]{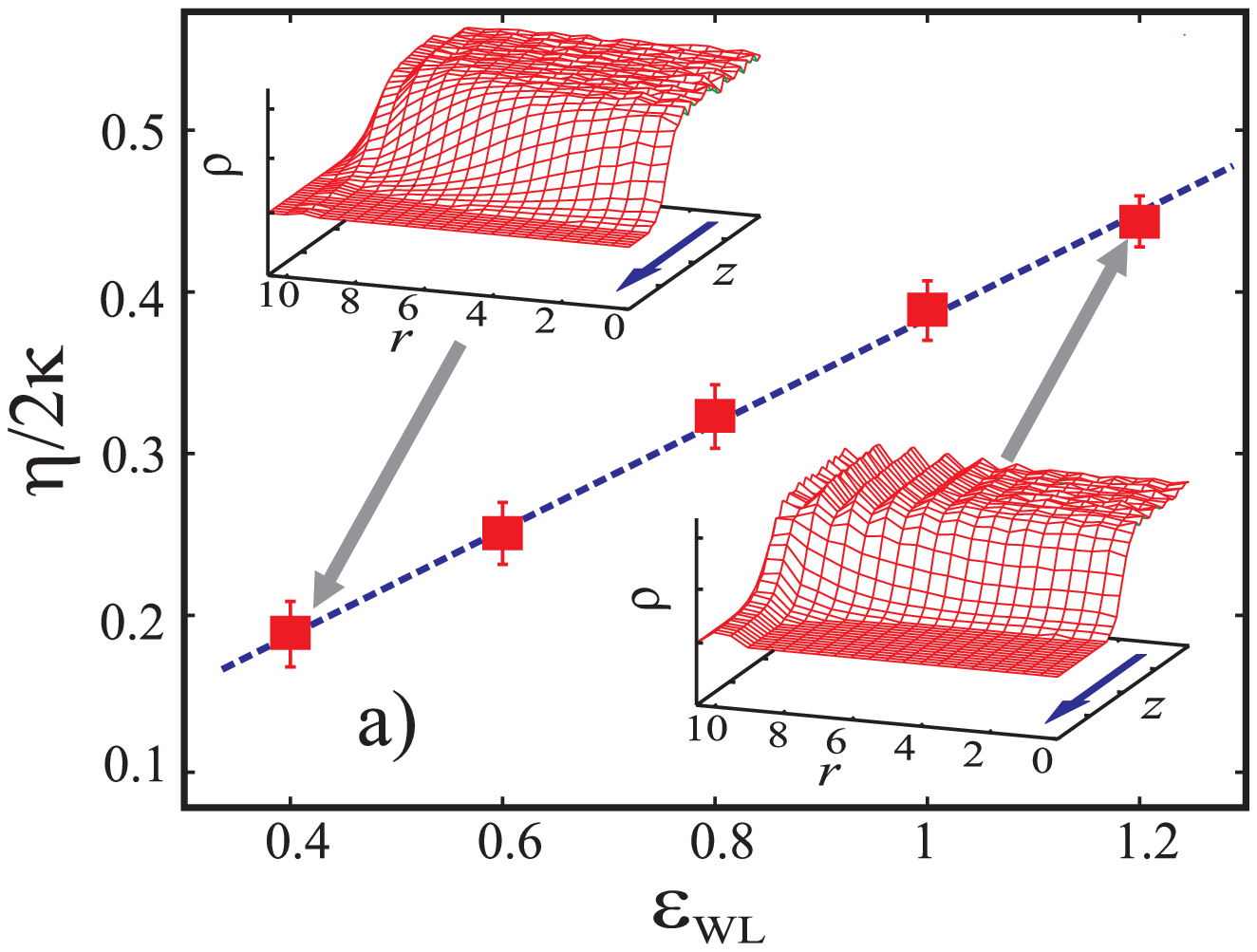}
\includegraphics[scale=0.6]{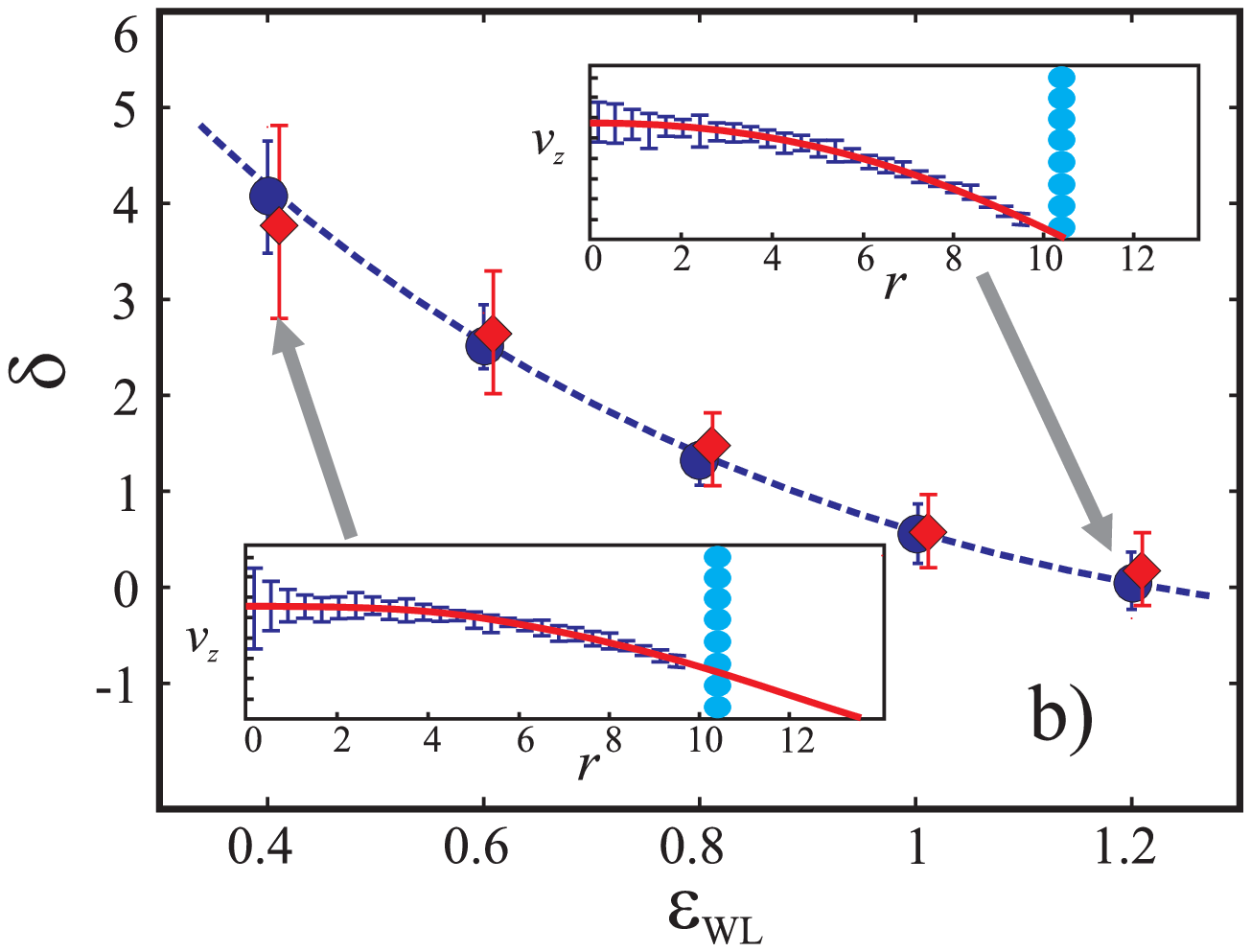}
\caption{(a) "Friction coefficient" $\eta/2\kappa$ plotted vs. $\epsilon_{WL}$.
The dashed line is a fit to a straight line. Insets show $3d$ density profiles
at the meniscus position for a nonwettable fluid (upper corner, left), and a
wettable one (lower corner, right).The wall position is at $r=11$ and the tube
axis - at $r=0$. Arrows indicate the direction of flow. Note the difference in
meniscus curvature between the two cases.
(b) Slip length $\delta=\sqrt{8\kappa}-R$ vs. wall-liquid interaction strength
$\epsilon_{WL}$ where $\kappa$ is determined from Eq.\ref{eq4}, full circles,
the dashed line being just a guide for the eye. Diamonds are derived from
extrapolating the radial distribution of axial velocity $v_Z$ in the meniscus
wake - see insets for $\epsilon_{WL}=0.4$ - bottom, left, and
$\epsilon_{WL}=1.2$ - top, right.
} \label{slip}
\end{figure}

Fig.~\ref{slip} shows
that also the "friction coefficient" of the imbibition (per unit length)
$\eta/2\kappa$ strongly depends on the wettability of the pore wall. The
computer experiment has the bonus that it yields insight into the behavior of
the system on the nanoscale in arbitrary detail. This is demonstrated by the
density profiles of the moving meniscus, shown for $\epsilon_{WL}=0.4$ and
$\epsilon_{WL}=1.2$, respectively. While no layering  of the fluid is observed
in the case of nonwetting fluids, $\epsilon_{WL}=0.4$, for a wettable wall the
profile for $\epsilon_{WL}=1.2$ indicates significant density oscillations in
the vicinity of the wall, i.e. fluid "layering"~\cite{8}.

Since the shear viscosity $\eta$ has been  determined independently for our
system~\cite{13}, $\eta = 6.34 \pm 0.15$, the ratio $\eta/2\kappa$ is readily
converted into an estimate for the slip length $\delta$ (Fig.~\ref{slip}b). The
gradual decrease of $\delta$ with growing wettability of the wall
$\epsilon_{WL}$ is clearly demonstrated.

In conclusion, we have modelled a possible and simple experimental  set-up
(Fig.~\ref{snap}) by computer simulation and provided a theoretical framework,
by slightly extending the Lucas-Washburn approach to include external pressure.
As demonstrated, this allows a consistent analysis of resulting data. Such an
analysis yields information on the Laplace pressure (if the pore radius is
known, the contact angle then can be estimated) as well as the permeability (and
hence the slip length).  The consistency of our description has been tested by
simulations where all these quantities were obtained independently. Thus we have
provided a framework which should be a useful guide for both experimental work
on capillary filling and further related simulations.

\paragraph*{Acknowledgments}: One of us (D.~D.) receives support from the
Max-Planck Institute of Polymer Research via the MPG Gesellschaft, another (A.
M.) received partial support from the Deutsche Forschungsgemeinschaft (DFG)
under project no 436BUL113/130. A.~M. and D.~D. appreciate support by the
project "INFLUS", NMP-031980 of the VI-th FW programme of the EC. We are
grateful to P. Huber and F. Mugele for stimulating discussions.

\end{document}